\documentclass[a4paper]{jpconf}
\usepackage{graphicx}
\begin{document}

\def\be{\begin{equation}}
\def\ee{\end{equation}}
\def\al{\alpha}
\def\bea{\begin{eqnarray}}
\def\eea{\end{eqnarray}}

\title{Graphene: QFT in curved spacetimes close to experiments}

\author{Alfredo Iorio}

\address{Faculty of Mathematics and Physics, Charles University in Prague, V Hole\v{s}ovi\v{c}k\'ach 2, 18000 Prague 8, Czech Republic}

\ead{iorio@mff.cuni.cz}

\begin{abstract}
A recently proposed step-by-step procedure, to merge the low-energy physics of the $\pi$-bonds electrons of graphene, and quantum field theory on curved spacetimes, is recalled. The last step there is the proposal of an experiment to test a Hawking-Unruh effect, emerging from the model, that manifests itself as an exact (within the model) prediction for the electronic local density of states, in the ideal case of the graphene membrane shaped as a Beltrami pseudosphere. A discussion about one particular attempt to experimentally test the model on molecular graphene is presented, and it is taken as an excuse to solve some basic issues that will help future experiments. In particular, it is stated that the effect should be visible on generic surfaces of constant negative Gaussian curvature, that are infinite in number.
\end{abstract}

\section{Introduction}

Recently (see \cite{iorio}, \cite{ioriolambiase} and, for a review, \cite{iorioLor}), in order to realize in practise a quantum Dirac field theory on curved spacetimes, there has been interest in exploiting certain special features of the low-energy physics of the $\pi$-bonds electrons of graphene, Weyl symmetry \cite{lor} being among them \cite{iorio}. The clearest signature of quantum field theory (QFT) on curved spacetime is the Hawking effect \cite{birrellanddavies}. In \cite{ioriolambiase} the best conditions for the simplest realization of the latter on graphene were proposed.

\begin{figure}
 \centering
 \resizebox{0.6\textwidth}{!}{%
  \includegraphics{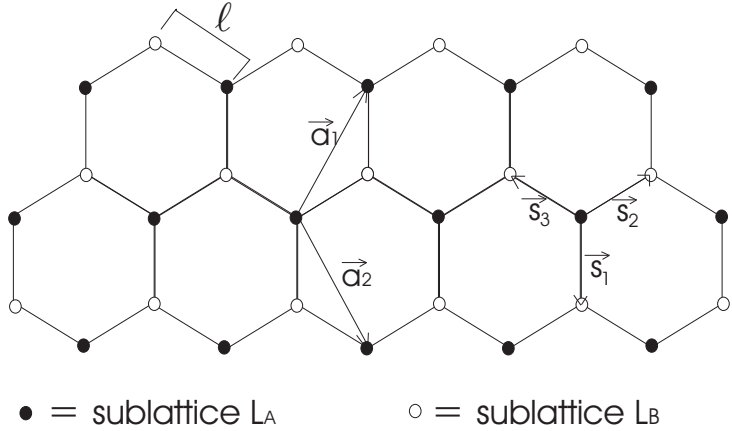}
  }
  \caption{\label{honeycomb} The honeycomb graphene lattice. The basis vectors, $\vec{a}_1 = \frac{\ell}{2} (\sqrt{3}, 3)$, $\vec{a}_2 = \frac{\ell}{2} (\sqrt{3}, - 3)$, link sites of the sublattice $L_A$, but cannot reach the sublattice $L_B$. For the latter sites the extra vectors, $\vec{s}_1 = \ell (0, - 1)$, $\vec{s}_2 = \frac{\ell}{2} (\sqrt{3}, 1)$, $\vec{s}_3 = \frac{\ell}{2} (-\sqrt{3}, 1)$, are necessary.}
\end{figure}

\begin{figure}[h]
\begin{minipage}{16pc}
\includegraphics[width=16pc]{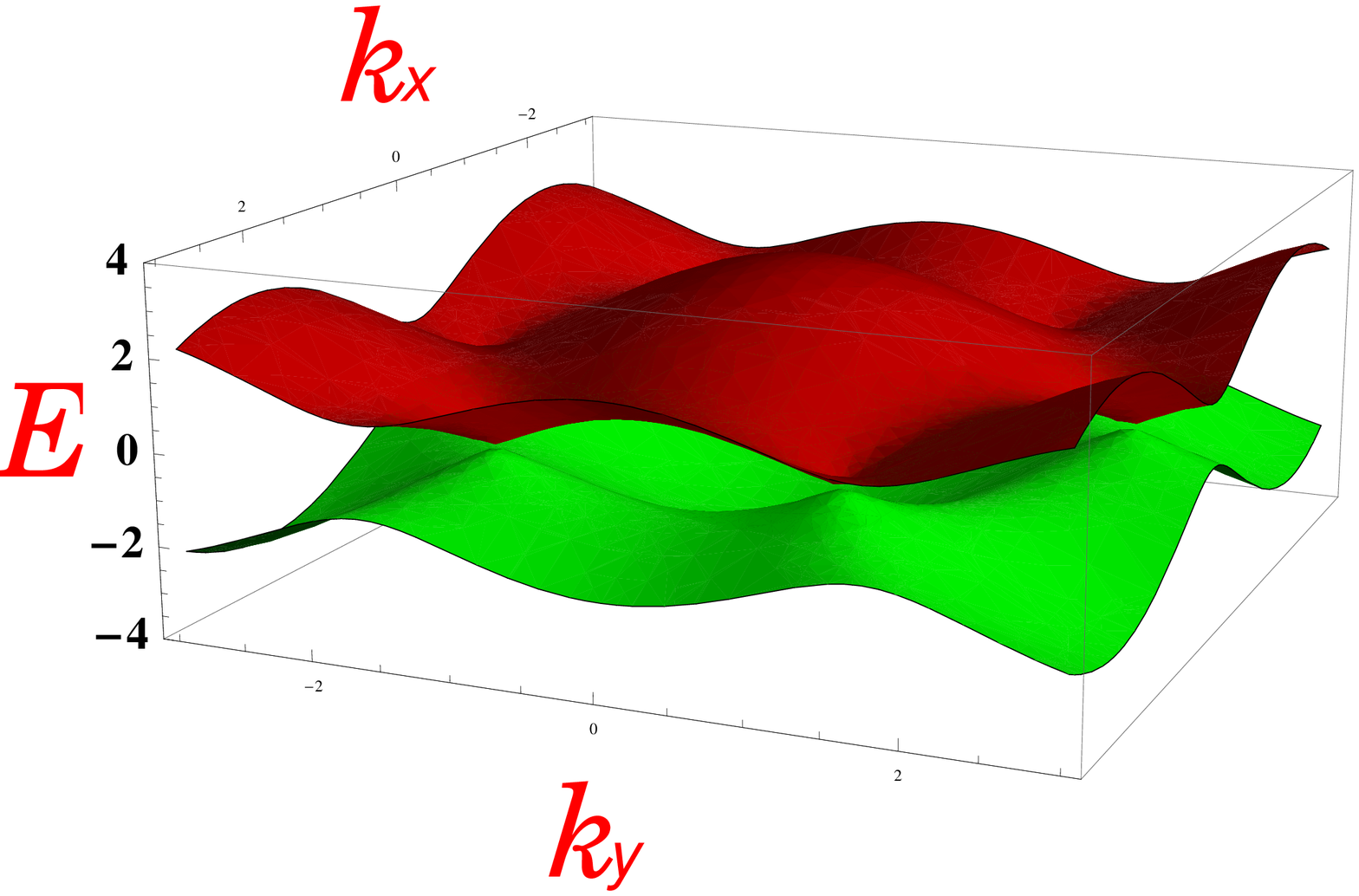}
\caption{\label{full}The dispersion relations of graphene, $E(k_x, k_y)$. The energy $E$ is in units of the hopping, $t \simeq 2.8$~eV. The conductivity band (red) and the valence band (green) touch in six {\it points} (where $E=0$), reflecting, in the Brillouin zone, the hexagonal symmetry of the lattice. Only the two points $\vec{k}^D_\pm $ are inequivalent.}
\end{minipage}\hspace{2pc}%
\begin{minipage}{20pc}
\includegraphics[width=20pc]{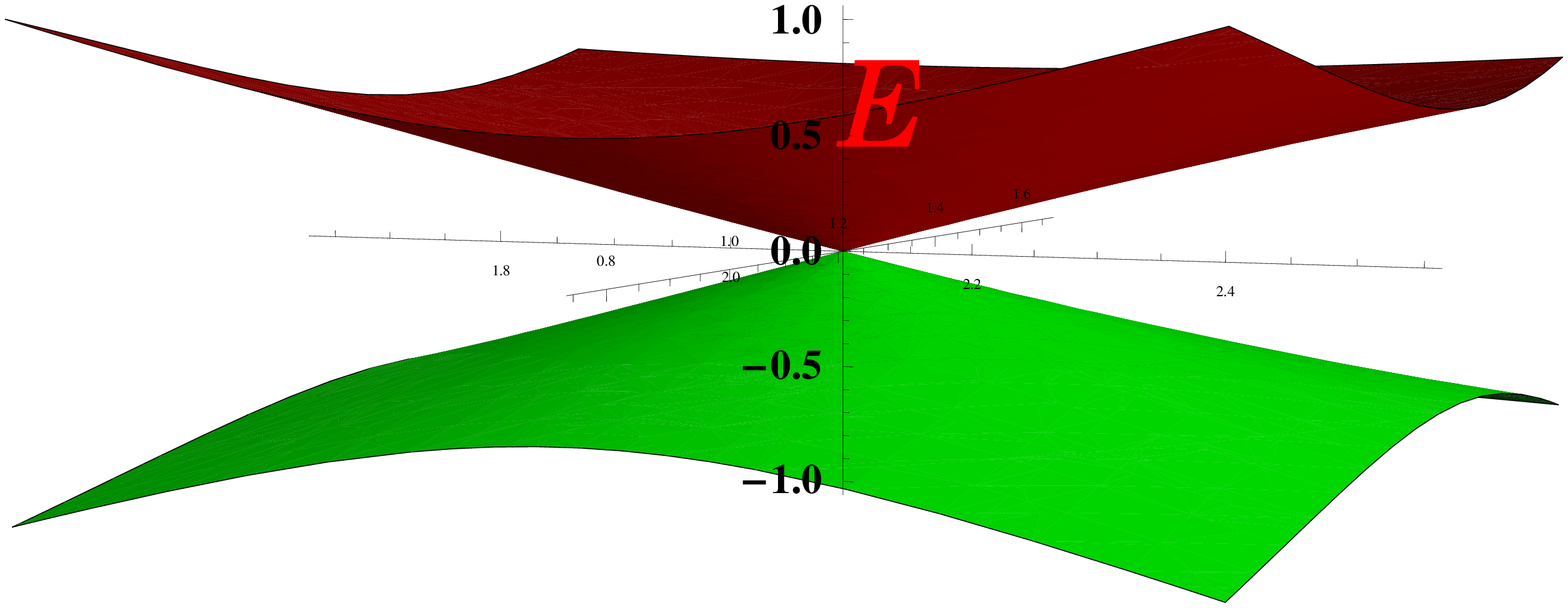}
\caption{\label{linear}The linear dispersion relations near one of the Dirac points, showing the typical behavior of a relativistic-like system (the ``$v_F$-light-cone''). Note that the energy range is of the order of one unit of the hopping energy $t$, i.e., here $E \in [- 3 {\rm eV}, + 3 {\rm eV}]$. Within this range, the continuum field approximation is tenable.}
\end{minipage}
\end{figure}

On one side, we have graphene, a one-atom-thick membrane of carbon atoms arranged in a hexagonal lattice, see Fig.~\ref{honeycomb}, first discussed theoretically in \cite{wallace, semenoff}, then found experimentally in \cite{geimnovoselovFIRST}.
Its very peculiar transport properties are summarized by the following low-energy tight-binding Hamiltonian, see, e.g., \cite{pacoreview2009}, valid near the Fermi {\it points} $\vec{k}^D_\pm = (\pm \frac{4 \pi}{3 \sqrt{3}}, 0)$ in the Brillouin zone, see Figs. \ref{full} and \ref{linear},
\be
 H  =  \frac{3 t \ell}{2} \sum_{\vec{p}} \left(\psi_+^\dagger \vec{\sigma} \cdot \vec{p} \; \psi_+
 + \psi_-^\dagger \vec{\sigma}^* \cdot \vec{p} \; \psi_- \right) =  - i \hbar v_F \int d^2 x \left( \psi_+^\dagger \vec{\sigma} \cdot \vec{\partial} \; \psi_+ + \psi_-^\dagger \vec{\sigma}^* \cdot \vec{\partial} \; \psi_- \right) \;, \label{HGraphene}
\ee
where $t \simeq 2.8$eV, is the electron energy for hopping to the nearest-neighbor, $\ell \simeq 1.4$\AA $\,$ is the carbon to carbon distance in the hexagonal honeycomb lattice, $v_F \equiv 3 t \ell / (2 \hbar) \simeq c/300 \simeq 10^6 {\rm m}/{\rm s}$, and with $\vec{\sigma} \equiv (\sigma_1, \sigma_2)$, $\vec{\sigma}^* \equiv (-\sigma_1, \sigma_2)$, $\sigma_i$ the Pauli matrices and $\psi_\pm \equiv \left( a_\pm , b_\pm \right)^{\rm T}$ Dirac spinors, where $a_\pm \equiv a(\vec{k}^D_\pm)$ ($b_\pm \equiv b(\vec{k}^D_\pm)$) is the annihilation operator for the electron in the $L_A$ ($L_B$) sublattice, evaluated at the Fermi points.

On the other side we have QFT in curved spacetime. This is by now an old theory \cite{birrellanddavies}, with no ultimate experimental evidence, hence with many open problems, and interesting issues to investigate in relation to the practical implementation of a fundamental theory. The open questions range from the structure of the quantum vacuum, and the related role of the inequivalent quantizations (think, for instance, to the {\it querelle} between the young and the old Hawking, about the information loss in the presence of a black hole), to the very meaning of particles in a curved spacetime. There are then open questions relative to the Hawking phenomenon itself, predicted but to-date impossible to detect from astrophysical sources, and the associated black-hole thermodynamics. These questions, although old, are all still on the table and permeate the contemporary research on high energy theory, as, for instance, one direction pursued these days is to see gravity, and the spacetime itself, as an emergent phenomenon.

Graphene appears to be a system where many of the issues above could be experimentally addressed. In the following, I will propose arguments and experiments in favor of this.

\section{The step-by-step merging}

The preliminary step to start the merging is to consider the dimensionality of the problem, that is (2+1) dimensions. I will take this duly into account. For instance, it is by now well known that many features of (3+1) dimensions need to be changed on the gravity side \cite{thooft}, \cite{djt}, \cite{cs3}, \cite{carlip}, that a black-hole is possible \cite{btz}, and that certain features, on the matter side, such as the ``statistical swapping'' \cite{takagi}, can take place.

$\bullet$ The $1^{\rm st}$ step is to take time on board, hence we move from the Hamiltonian (\ref{HGraphene}) to the corresponding action
\be
    A =  i \hbar v_F \int d^3 x \bar{\psi} \; \gamma^a \partial_a \; \psi \;,
\ee
where $a=0,1,2$, $\gamma^0 = \sigma_3$, $\gamma^1 = i \sigma_2$, $\gamma^2 = - i \sigma_1$, and, since we do not consider phenomena mixing the two Fermi points, we focus on one point only, e.g., $\psi \equiv \psi_+$. When the graphene sheet is curved, and within certain limits \cite{iorioLor}, the action to consider is
\begin{equation}
{\cal A} = i \hbar v_F \int d^3 x \sqrt{g} \; \bar{\psi} \gamma^\mu \nabla_\mu \psi \label{gigino} \;,
\end{equation}
where $\nabla_\mu = \partial_\mu + \Omega_\mu$, and $\Omega_\mu \equiv \frac{1}{2} \omega_\mu^{\; a b} J_{a b}$ is the gauge field able to take into account intrinsic curvature. Here $\omega_\mu^{\; a b}$ is the spin connection, and $J_{a b}$ are the generators of local Lorentz transformations, with $\mu, ...$ responding to diffeomorphisms, and $a, b, ...$ flat indices. Within the limits indicated in \cite{iorioLor} and \cite{ioriolambiase2} (essentially, in the absence of torsion and of strong scatters \cite{peres}, and for small curvatures, i.e. large radius of curvature w.r.t. the lattice, $r > \ell$), this action well describes the physics of the very long wavelength $\pi$ electrons on curved graphene sheets. The only effects it needs to take care of are the ones induced by intrinsic curvature, i.e., \textit{inelastic} effects that we ascribe to disclination defects \cite{KatVol92}.

$\bullet$ The $2^{\rm nd}$ step is to exploit the absence of mass in the action (\ref{gigino}), that, in the case in point of the Dirac field, guarantees full local Weyl symmetry \cite{lor}. In the case of conformally flat metrics, $g_{\mu \nu} = \phi^2 \eta_{\mu \nu}$, Weyl symmetry gives
\be
i \int d^3 x \sqrt{g} \bar{\psi} \gamma^\mu \nabla_\mu \psi = i \int d^3 x \bar{\psi}' \gamma^a \partial_a \psi'
\ee
where, under Weyl's transformations, $\psi = \phi^{-1} \psi'$. Namely, the flat space theory is classically equal to a curved space theory.
We have to consider quantum effects, though. As we are in a odd-dimensional spacetime, this does not refer to quantum trace anomalies, absent here \cite{chr}. What we refer to here (see \cite{iorio, TFDBH}) is that one can, essentially, do two type of things: (a) perform measurements in a frame where the quantum effects of curvature disappear, e.g.,
\be
 S (x_1, x_2) \equiv \langle 0| \psi  (x_1) \bar{\psi} (x_2) |0 \rangle = \;
    '\!\langle 0| \psi ' (x_1) \bar{\psi}' (x_2) |0 \rangle '  = S ' (x_1, x_2) \;,
\ee
where $S (x_1, x_2)$ is the two-point function in the curved case, and $S' (x_1, x_2)$ is the two-point function in the flat case; or (b) perform the measurements in a frame where, due to Weyl symmetry, the effects are seen as a simple modification of the flat case (conformal triviality \cite{birrellanddavies})
\be \label{green}
S (x_1, x_2) \equiv  \; '\!\langle 0| \psi  (x_1) \bar{\psi} (x_2) |0 \rangle '
    = \phi^{-1} (x_1) \phi^{-1} (x_2) \; S ' (x_1, x_2) \;.
\ee
In our case, due to the obvious difficulty to realize a frame where the measuring apparatus is in a gravitating field whose spacetime coincides with the graphene curved spacetime, the second choice will be the right one.

$\bullet$ The $3^{\rm rd}$ step is to stick to a metric that is the most easy to realize in a lab, namely
\be
g^{(2+1)}_{\mu \nu} (x,y) = \left(\begin{array}{cc} 1 &  \\  & g^{(2)}_{\alpha \beta} (x,y)  \end{array} \right) \;,
\ee
i.e., with all the curvature in the spatial part $g^{(2)}_{\alpha \beta}$, and no time-dependence. Can this be conformally flat, rich and still interesting? To answer all these questions we need to solve \cite{iorio, cs3}
\be
    C_{\mu \nu} = \epsilon_{\mu \lambda \kappa} \nabla^\lambda {R^{(2+1)}}^\kappa_\nu
    + \epsilon_{\nu \lambda \kappa} \nabla^\lambda {R^{(2+1)}}^\kappa_\mu = 0 \;.
\ee
Since all surfaces of constant Gaussian curvature $\cal K$ solve this equation \cite{iorio}, they give rise to conformally flat (2+1)-dimensional spacetimes $g^{(2+1)}_{\mu \nu}$. Hence the answer to the above questions is yes: by simply curving the graphene sheet one can have rich enough situations (surfaces of constant ${\cal K}$) implying a conformally flat spacetime.

$\bullet$ The $4^{\rm th}$ step. The latter result is intrinsic, but to implement Weyl symmetry we need the explicit conformal factor, and this might mean to change the frame
\be
q^\mu \equiv (t, x,y) \rightarrow {\cal Q}^\mu \equiv (T,X,Y)
\ee
so that
\be
g^{(2+1)}_{\mu \nu} ({\cal Q}) = \phi^2 ({\cal Q}) \; g^{{\rm flat}}_{\mu \nu} ({\cal Q}) \;.
\ee
This can be technically difficult (a system of 6 partial differential equations), but the most important here are the physical meaning, global predictability, and practical feasibility of this coordinate frame.

The surfaces of ${\cal K} = + 1/r^2$= const. can be described by the spatial line element \cite{eisenhart}
\be
d\ell^2 = du^2 + c^2 \cos^2 (u/r) dv^2
\ee
where $u$ is meridian coordinate, and $v$ the parallel coordinate, and with the real constant $c$ such that $c = r$ (sphere), $c < r$, $c >r$. These
${\cal Q}^\mu$ are difficult to find.

\begin{figure}
\centering
\resizebox{0.5\textwidth}{!}{%
\includegraphics{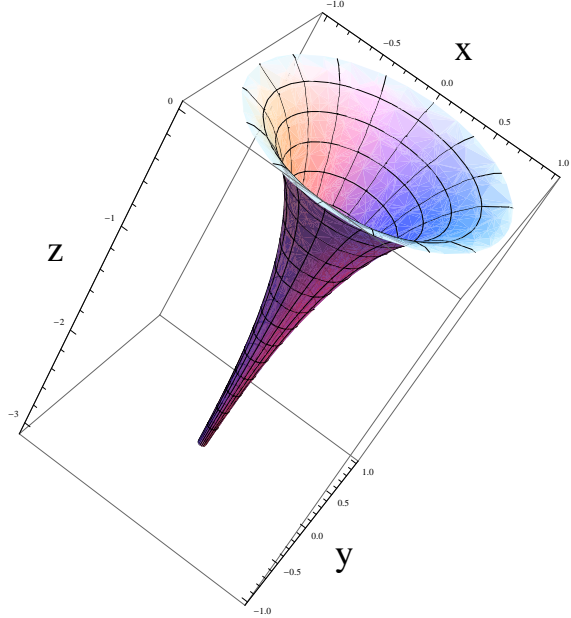}
}
\caption{The ${\bf R}^3$ coordinates of the Beltrami pseudosphere, in the canonical form, are \cite{spivak} $x(u,v) = R(u) \cos v$, $y(u,v) = R(u) \sin v$, $z(u) = r (\sqrt{1 - R^2(u)/r^2} - {\rm arctanh}\sqrt{1 - R^2(u)/r^2})$, with $R(u) =  c \, e^{u/r}$, $c > 0$ and $r = \sqrt{-{\cal K}^{-1}} > 0$ where ${\cal K}$ is the constant negative Gaussian curvature.  We can choose $c=r$, thus $R(u) \in [0,r]$ as $u \in [-\infty, 0]$. The surface is not defined for $R>r$ ($z$ becomes imaginary), an instance of the Hilbert theorem. In the plot $r=1$ and $u \in [-3.37, 0]$, $v\in [0, 2\pi]$.} \label{Beltrami}
\end{figure}

Interesting objects, like, e.g., the black-hole of \cite{btz}, are related to spacetimes of negative constant curvature. For the surfaces of ${\cal K} = - 1/r^2$= const. the following theorem by Hilbert holds: {\it No non-singular surface of constant ${\cal K} = - 1/r^2$ can be embedded in ${\bf R}^3$}. If we can live (and we can) with singularities, such surfaces are many more than the positive curvature counterparts (there are, actually, an infinite number of them). Among them there are the surfaces of revolution, called pseudospheres, and they can be described by the spatial line element \cite{eisenhart}
\be \label{pseudospheres}
d\ell^2 = du^2 + \left( c_1^2 \cosh^2 (u/r) + c_2^2 \sinh^2 (u/r) \right) dv^2
\ee
with $c_1, c_2 \in {\bf R}$. For all these surfaces (and also for those not included in the above line element)
\be \label{graphenemetric}
ds^2_{\rm graphene}=dt^2-\frac{r^2}{{\tilde y}^2}(d{\tilde x}^2+d{\tilde y}^2)
= \frac{r^2}{{\tilde y}^2}\left[\frac{{\tilde y}^2}{r^2}dt^2-d{\tilde x}^2-d{\tilde y}^2\right]
\ee
with ${\tilde x}, {\tilde y}$ the abstract coordinates of upper-half plane model of the Lobachevsky geometry (see \cite{ioriolambiase2} for the case in point of graphene, and \cite{penrose} for a general introduction to the subject). To make sense in a laboratory, these coordinates need to be expressed in terms of measurable coordinates, say the $(u, v)$ coordinates above, and this amounts to choose the surface. The line element in square bracket is flat, but this is not yet the solution of our problem, because we need to see, case by case, to which physical ${\cal Q}^\mu$ the coordinates $(t, {\tilde x}, {\tilde y})$ correspond to.

As a counter-example, consider the case of the hyperbolic pseudosphere ($c_1 \neq 0$ and $c_2 = 0$ in (\ref{pseudospheres}))
\be
\tilde{x} = e^{c v/r} \tanh (u/r) \;, \quad \tilde{y} = e^{c v/r} / \cosh (u/r) \;.
\ee
Since the conformal factor $r^2/{\tilde y}^2$ in (\ref{graphenemetric}) is multivalued ($v$ is an angle), this solution as no global predictability in $q^\mu = (t,u,v)$, hence one must look for a different frame where this does not happen, but the frame in point might be difficult to realize in practise (and surely it is not the lab frame $q^\mu$, where all that needs to be done is to curve the surface).

It is the Beltrami pseudosphere (see Fig.~\ref{Beltrami})  that solves the problem
\be
\tilde{x} = v/r \;, \quad \tilde{y} = e^{-u/r}/r
\ee
i.e. $r^2 / \tilde{y}^2$ makes sense over the whole surface/spacetime. Hence, the physical ${\cal Q}^\mu$ are the lab's $q^\mu$. A physically doable $g^{\rm graphene}_{\mu \nu} = \phi^2  \; g^{\rm flat}_{\mu \nu}$ is already there in the frame $q^\mu$:
\be
ds^2_{\rm graphene} \equiv ds_{(B)}^2 = e^{2u/r} \left[e^{-2u/r}(dt^2-du^2)-r^2dv^2\right] \;.
\ee
Now we can say that the line element in square brackets is the Rindler line element, $ds_{(R)}^2$, and exploit this fact in our quest for measurable effects of the QFT in curved space interpretation of the physics of graphene $\pi$ electrons.

\begin{figure}
\centering
\resizebox{0.8\textwidth}{!}{%
\includegraphics{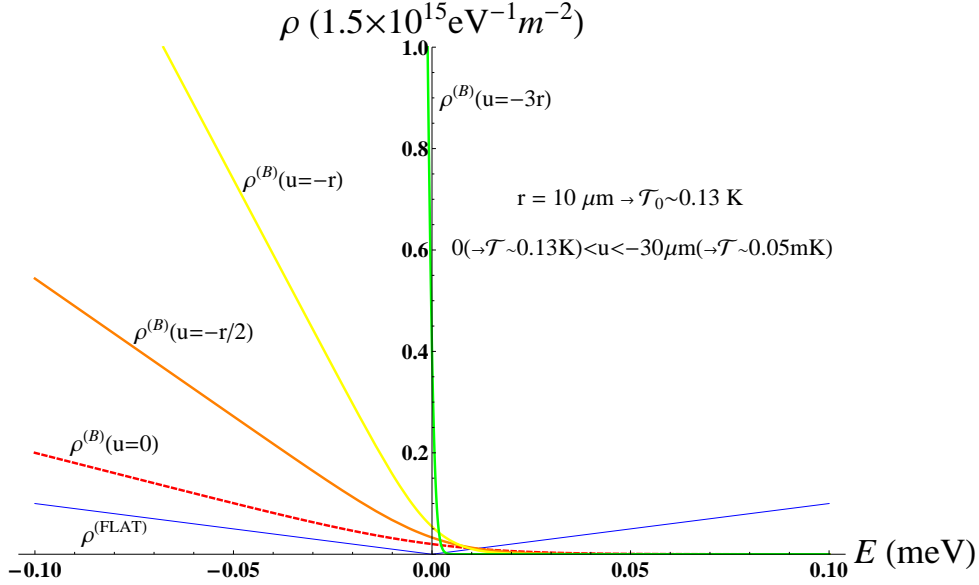}
  }
\caption{\label{linecut} Series of $dI/dV$ spectra (``line-cut'') from an STM. The model applies to energies below the natural scale, $E^* = \hbar v_F/r \simeq 6.6 \times 10^{-7} {\rm eV} \times [r \; {\rm mm}]^{-1}$, where $[r \; {\rm mm}]$ is the numerical value of $r$ measured in mm. In this plot $E \in [-0.1,0.1]$meV, corresponding to $r = 10\mu$m. The flat LDOS is shown in blue. $u \in ]0, -3r]$, $u=0$ (red dashed curve) corresponds to the singular boundary of the pseudosphere, beyond the upper limit $u=-3r$ (green curve) the pseudosphere is too sharp. To each spectrum corresponds a $u$-dependent temperature $\cal T$ while the constant temperature, same for the whole line-cut, is ${\cal T}_0 = e^{-u/r}{\cal T} \sim 0.13$K.}
\end{figure}

$\bullet$ The $5^{\rm th}$ step consists in implementing the Unruh requests on this setting. In particular it is important the choice of the quantum vacuum of reference, taken here to be the Minkowski vacuum $|0\rangle' \equiv |0\rangle_M$, thus the Green's function of interest are of the form (see (\ref{green}) and \cite{iorio, ioriolambiase, ioriolambiase2})
\be \label{green2}
S^B (x_1, x_2) \equiv \langle 0_M | \psi^B  (x_1) \bar{\psi}^B (x_2) |0_M \rangle
    = \phi^{-1} (x_1) \phi^{-1} (x_2) \;  \langle 0_M | \psi^R  (x_1) \bar{\psi}^R (x_2) |0_M \rangle \;,
\ee
where $B$ stands for ``Beltrami'', $R$ for ''Rindler'', and the Green's function on the right side is $S^R$, the customary two-point function used in the computation of the Unruh effect. The power spectrum is customarily computed by taking $S^R$ at the same point in space, $\vec{q}$, and at two different times, with $t_2 - t_1 = e^{u/r} \tau $, and $\tau$ the proper time (the factor $e^{u/r}$ is related here to the proper acceleration, and needs be constant, to fit the requests for the standard Unruh effect). From this, by Fourier transforming on the time variable, we get \cite{takagi}
\be
F^{R}(\omega, \vec{q}) \equiv
\frac{1}{2} {\rm Tr} \left[\gamma^0\int_{-\infty}^{+\infty}d\tau e^{-i\omega \tau} S^{R}(\tau, \vec{q}, \vec{q})\right] \;,
\ee
and it coincides here with the (not yet physical) electronic local density of states (LDOS) \cite{pacoreview2009}
\be
    \rho^{R}(\omega, \vec{q}) \equiv \frac{g}{\pi} F^{R}(\omega, \vec{q}) \;.
\ee
As this is a massless case, the result of the computation is exact
\be
    F^{(R)}(\omega)=\frac{1}{2}\, \frac{\omega}{e^{\omega/{\cal T}}-1}
\ee
where ${\cal T}\equiv \alpha(u) / (2\pi) =  e^{u/r} / (2\pi r)$. We are now in the position to make a verifiable prediction, by going back to the physical spacetime. Due to the Weyl symmetry, this is easy (see (\ref{green2}))
\be
\rho^{B}(\omega, \vec{q}) = \phi^{-2}(\vec{q}) \rho^{R}(\omega, \vec{q}) \;.
\ee
Thus, the LDOS predicted for a graphene sheet shaped as a Beltrami pseudosphere is \cite{ioriolambiase}
\be \label{THEformula}
 \rho^{B}(E, u, r)= \frac{4}{\pi} \frac{1}{(\hbar v_F)^2} \frac{E \; e^{-2 u /r}}{\exp{\left[ E / (k_B {\cal T}_0 e^{ u/r}) \right]}-1} \;,
\ee
i.e., a Hawking phenomenon with Hawking temperature
\be
{\cal T}_0 \equiv \frac{1}{2 \pi} \; \frac{\hbar v_F}{k_B \, r} \;.
\ee
In Fig.~\ref{linecut} a plot of what we expect from a dedicated experiment with a Scanning Tunneling Microscope \cite{ioriolambiase}.

\section{Paving the way to future experimental set-ups}

Lately, I have been interacting with experimentalists from the Manoharan Lab of Stanford University\cite{manoharan}, who asked whether it is possible to see effects, of the kind described above, on their ``molecular graphene'' \cite{artgraph}, that is a planar hexagonal lattice of carbon monoxide, CO, molecules that mimics graphene. Unfortunately, there were severe limitations in their experimental settings. The first is the flat setting, the second is the small number of points, the third is that molecular graphene is a simulator of the real thing that, although a very good one to test certain properties, it has basic parameters (e.g., the neighbor-to-neighbor distance, $\ell$) that make our task more difficult. 

To write things in a list, what I had to address are the following issues:
\begin{enumerate}
  \item How to tile a Beltrami pseudosphere?
  \item Would a ``planar development'' (the result of mapping the surface on a plane, hence with curvature traded for strain), still captures the features necessary for the Hawking-Unruh phenomenon to happen? Which one, among the infinitely many, is the best?
  \item Is the effect still there for a {\it generic surface} of negative constant $\cal K$? If yes ...
  \item ... can we solve issues (i) and (ii) for the given surface?
  \item Can we spot any Hawking-Unruh in a planar configuration of molecular graphene, that mimics the planar development of a {\it portion} of a generic graphene surface of (not even necessarily constant) negative $\cal K$?
\end{enumerate}
My conclusions for the last point (the best meeting point between Manoharan Lab's and this model's needs), are negative, as I will explain below. This analysis gave extra arguments in favor of a dedicated experiment, with a Beltrami pseudosphere, as the best way to test this model. Nonetheless, the attempt was worth the effort. First, I learned that point (iii) has a positive solution. Thus, the realization of a generic graphene surface of constant negative Gaussian curvature, although not being the best possible setting, should be enough to test a modified formula (\ref{THEformula}), at least in an approximate way. On this I will briefly comment below, and will write in detail elsewhere. Second, for whatever dedicated experiment one would like to carry on, matters like the issue (i) in the list are crucial. In the following, I will address in detail this latter issue, and will also briefly comment about the other issues.

Therefore, let me focus on the geometric and topological constraints on building a Beltrami pseudosphere with the hexagonal graphene membrane. There are two issues here. The first is the infinite (non-compact) surface ($u\in [-\infty, 0]$). The second is how many heptagonal defects are necessary. Let me introduce, briefly, the matter by recalling how to tile a compact surface $\Sigma$ (see, e.g., \cite{ioriosen}). Suppose we want to tile $\Sigma$ with regular $s$-sided polygons assembled in such a way that each vertex is shared by 3 polygons, and each edge is shared by 2 polygons. If $n_s$ is the number of polygons used (there are, say, $n_5$ pentagons, $n_6$ hexagons, $n_7$ heptagons, etc.) to tile $\Sigma$, the resulting \textit{polyhedron} $P$ has $V_P = 1/3 \sum_n n_s s$ vertices, $E_P = 1/2 \sum_n n_s s$ edges, and $F_P = \sum_n n_s$ faces, giving for the Euler characteristic
\be \label{euler}
\chi (\Sigma) = V_P - E_P + F_P \;,
\ee
the following expression
\begin{equation}\label{ngons}
  \sum_n n_s (6 - s) = 6 \chi (\Sigma)\;.
\end{equation}
For the sphere, the Gauss-Bonnet theorem\cite{spivak}, that links the topological number, $\chi$, to the geometric object that is total curvature, gives
\be
K_{tot} \equiv \int {\cal K} \; d^2\mu = 2 \pi \chi \label{GBsimple} \;,
\ee
and, being ${\cal K} = 1 /r^2=$~constant
\be
\chi = \frac{\rm Area}{2 \pi r^2} \;, \label{GBsimpleconstGauss}
\ee
or, being Area~$= 4 \pi r^2$, we have $\chi = 2$. Hence, if we try to wrap the hexagonal lattice of graphene, to form a tiling of the sphere, we cannot match $\sum_n n_s (6 - s) = 12$, only with hexagons (whose number is not constrained by this argument), but need something else. Considering that the creation of each defect costs energy, the minimum number of defects we need are 12 pentagons, $\sum_n n_s (6 - s) = n_5 \; (6 - 5)  + n_6 \; (6 - 6) = 12$, with which we could make the smallest (and ``minimum energy'') tiling of the sphere, the dodecahedron ($F=12$, $V=20$, $E=30$). Adding $20$ hexagons to it, one makes a buckyball (fullerene). One can increase $n_6$ further (according to the rule $n_6 = 10 (T -1)$, with $T = m^2 + l^2 + m l = 0, 1, 3,4,7,9,...$, and excluding $T=0$), making ``inflated'' buckyballs. Once we trade the continuous sphere for its tiling polyhedron, the curvature resides in the vertices. For the sake of simplicity, let us focus on the icosahedron, dual ($F \leftrightarrow V$, $E$ the same) to the dodecahedron,
so that $K_{tot} = 12 \; K^{ico}_V = 4 \pi$, and $K^{ico}_V \equiv K_5$. The curvature associated to each pentagon is then
\be \label{curvpent}
K_5 = + \pi /3 \;,
\ee
while no curvature is associated to the hexagons, $K_6 = 0$. Of course, the same results can be obtained using the dodecahedron, although the argument becomes more involved. A direct way to obtain this result is with the {\it defect/excess angle} argument. The curvature at one vertex of the polyhedron is
\be \label{defeccangle}
K_V \equiv 2\pi - \sum_i \alpha_i \;,
\ee
where $\alpha_i$ is the angle between edges of each face at the vertex. For a pentagon (formed with equilateral triangles in the icosahedron), $\alpha_i = \pi /3$, and $i = 1, ..., 5$, hence, $K_V = 2\pi - 5 \pi /3 = + \pi /3$ (defect angle). The same argument, for an hexagon gives $K_V = 2\pi - 6 \pi /3 = 0$.

Notice that the radius of curvature, $r$, does not enter these formulae, hence the curvature associated to each defect is fixed. If we want to increase the curvature of the sphere (make $r$ smaller), we do not have to increase the {\it number} of defects, but their {\it density} (i.e., we have to decrease the number of hexagons).

From the elastic point of view, it has been shown that, for the sphere, the creation of {\it heptagonal} defects is energetically favored in certain conditions \cite{bowick}, but this does not violate the Euler-Poincar\'e result (\ref{euler}), provided they come in pairs with pentagonal defects, $n_5 - n_7 = 12$, i.e. $n_5 = 12 + x$ and $n_7 = x$, and that is indeed the case (scars \cite{bowick}). The same argument descending from the Gauss-Bonnet theorem (\ref{GBsimple}) (and the excess angle argument of (\ref{defeccangle})) gives
\be \label{curvhept}
K_7 = - \pi /3 \;,
\ee
as the curvature associated to each one of these defects. Thus, when we want to build (tiling of) surfaces of negative $\cal K$ we need heptagonal defects.

If we want to wrap the graphene membrane into a Beltrami pseudosphere, we first need to consider that the issue of an infinite (non-compact) surface is actually minor, from a practical (lab) point of view. One can truncate the Beltrami, without changing the topology, and still has a surface of constant negative curvature. (Notice, also, that the elliptic and hyperbolic pseudospheres are finite surfaces, so, for them, this problem is absent).
So one can think of tiling the surface in the standard way illustrated above. The issue here is the number and types of defects necessary, and, in turn, how the Euler-Poincar\'e and Gauss-Bonnet theorems apply here, as they have been conceived for compact surfaces.

Having the topology of an annulus, the Euler number for Beltrami is \cite{gibbonseuler}
\be
\chi_{\rm Beltrami} = 0 \;.
\ee
(The same is true for the other two pseudospheres, not clear is the matter for more exotic pseudospheres \cite{gibbonseuler}). If one blindly applies (\ref{euler}), $V - E + F = 0 \to \sum (6 - s) n_s = 0$, the implications do not make sense. Would that mean that $s=6$? So I can tile with hexagons all over? But that is true for the plane, the
cylinder, or the cone, not for the pseudosphere. Would that mean that $n_7 = n_5$? But then I can satisfy the constraint with even 1 heptagon and 1 pentagon, and this makes no sense either. The unsolved (and intriguing) problem here is that, as far as I know, it is not known the negative curvature counterpart of the dodecahedron (or, of its dual, the icosahedron), the ``elementary'' polyhedron one obtains from the Euler-Poincar\'e for the sphere. There must be a fixed number of heptagons, that take into account the basic geometric properties, just like the 12 pentagons do for the sphere. The beauty and intricacies of the Lobachevsky geometry come in here, and make changes to the sphere story, and then introduce the singular boundaries, but the basic object should be there to give a unifying ``Platonic-Lobachevsky'' solid for all the infinite pseudospheres (not just the three that clearly have $\chi = 0$, but all the others). A partial solution to this problem, and in turn to our main problem (wrapping graphene to make a Beltrami), is to consider that the Gauss-Bonnet for surfaces with boundaries is \cite{spivak}
\be
\int_\Sigma {\cal K} d^2\mu + \int_{\partial \Sigma} {\cal K}_g dl = 2 \pi \chi \;, \label{GBfull}
\ee
with  ${\cal K}_g$ the geodesic curvature that measures how far is a curve from a geodesic (e.g., ${\cal K}_g = 0$ for the equator of a sphere). In (\ref{GBfull}) there are two parts, one for the bulk that is the total curvature, $K_{tot}$, and one for the boundaries. It is $K_{tot}$ that tells about the number of defects, the boundary term does not. Indeed, from ${\rm Area}  = 2 \pi r^2$, we have $K_{tot} =  - {\rm Area} / r^2 = - 2 \pi$, (the same is true for the other two pseudospheres), hence, using (\ref{curvhept}), $K_{tot} =  - 2 \pi  = n_7 \; K_7 = n_7 \; (- \pi / 3)$, that gives
\be
n_7 = 6 \;.
\ee
In other words, the Euler-Poincar\'e and the Gauss-Bonnet theorems ``decouple'' for pseudospheres, and we cannot apply (\ref{euler}) to compute the number of defects, but need the curvature arguments just given. That said, though, everything else is like for the sphere: the negative curvature counterpart of the icosahedron/dodecahedron, at least for the three classic pseudospheres, has 6 heptagons, arranged to form a non-compact ``polyhedron''(actually, I did not succeeded to use only heptagons, and a certain number of hexagons are necessary even for the basic object), and, elastic energy allowing, one can add pentagon-heptagon pairs. This is my solution of the issue (i) in the list above. Let me now face the other issues.

The procedure to map a surface onto a plane is called ``planar development'' \cite{spivak}. For the intrinsically flat surfaces, this can be done isometrically (that is why those surfaces are called ``developable''). For intrinsically curved surfaces, planar development comes at the price of deforming the distance among points (think of any map of the Earth), hence of trading curvature for strain, and there are infinitely many different ``planar versions'' of the curved surface, see, e.g., \cite{azariadis}. At first sight, the effects we are describing with (\ref{THEformula}), being due to intrinsic curvature, could never be seen on a planar development. In the case in point of the graphene membrane, though, there is some hope to still have some sort of signal of the Hawking-Unruh effect, although we can expect it to be substantially deformed by the planar development. It is so, because, in our model, the intrinsic curvature of the graphene membrane is encoded in the heptagonal defects, hence, retaining these defects in the developed Beltrami surface, in principle, could be the way out to test the effect on a flat sample. Unfortunately, the geometric constraints on the development, and the limits of the experimental set-up of the device of Manoharan Lab, conspired to make this test impossible. Let me explain in a bit more detail.

The actual values of various important quantities differ, for graphene and for molecular graphene (for the latter I rely on \cite{dominik}, see also \cite{manoharan} and \cite{artgraph}). In the table~\ref{secondtable}, I summarize the ones relevant for this discussion. There, I give two energy scales, $E_\ell \simeq \hbar v_F / \ell$ (below which the continuum Dirac field approximation holds), and $E_r \simeq \hbar v_F / r$ (below which the field theory in curved space time holds). The formula (\ref{THEformula}) has a chance to be tested only if $r >> \ell$, and it holds for energies below $E_r$. On this see \cite{iorioLor} and \cite{ioriolambiase, ioriolambiase2}. Essentially, what goes wrong with the Manoharan Lab set-up is that, even to just include all the necessary 6 heptagons, the limited number of points makes the density of defects (hence, the corresponding curvature) too close to the limit for the model to hold. Given the values of $v_F$ and $\ell$, and the maximum sensitivity of the STM machine of 1-10 meV, to a small enough curvature (big $r$) corresponds an energy too small to detect the effect in that range (see the entry for $r \sim 100 \ell$ in table~\ref{secondtable}, and consider that the electrons need to have an energy below the corresponding $E_r$).

\begin{table}[h]
\caption{Values of various important quantities, for graphene and for molecular graphene. Here $v_F = \frac{3}{2 \hbar} t \ell$, with $t$ hopping parameter, $\ell$ nearest-neighbor distance, and $E_\ell \simeq \hbar v_F / \ell$, $E_r \simeq \hbar v_F / r$.}
\centering
\begin{tabular}{|l|l|l|}
  \hline
   & Graphene & Molecular Graphene \\ \hline
  $\ell$ (\rm m) & $1.4 \times 10^{-10}$ & $20 \times 10^{-10}$ \\ \hline
  $v_F$ ({\rm m}/{\rm s})& $9 \times 10^5$ & $6.5 \times 10^5$ \\  \hline
  $E_\ell$ (\rm eV)& $4.2$ & $0.2$ \\  \hline
  $E_r (r \sim 100 \ell)$ (\rm meV) & $42$ & $2.1$ \\  \hline
  $E_r (r \sim 50 \ell)$ (\rm meV) & $84$ & 4 \\  \hline
  $E_r (r \sim 20 \ell)$ (\rm meV) & 200 & $10$ \\
  \hline
\end{tabular}
\label{secondtable}
\end{table}

Indeed, the number of defects per unit area is $d^{(7)} \equiv n_7/ A = 3/\pi \; r^{-2}$, as $n_7 = 6$ and $A = 2 \pi r^2$ for Beltrami. So the number of defects in the area $a$ of the surface is $n^a_7 = d^{(7)} \; a  = 3/\pi \; a / r^2$, hence
\be
r (n_7^a) = \sqrt{\frac{3 a}{\pi n_7^a}} \;.
\ee
The number of CO molecules is limited to 500, hence an estimate of the area is: 170 hexagons, each with area $3 \sqrt{3}/2 \ell^2$, giving $a \sim  450 \ell^2 \sim 9000{\AA}^2$. With these, $E_r \simeq \hbar v_F / r = \alpha^{-1} E_\ell = \alpha^{-1} \times 0.2 {\rm eV}$, where $\alpha = r/\ell$, all the quantities are for molecular graphene. In table~\ref{firsttable}, I report the estimate of the values of $r$ as a function of the defects present in the lattice, and the corresponding $E_r$s. It is clear from there that we are right at the limit of validity of the model, (see the entries for $n_7^a = 6$).

Furthermore, a planar development of the Beltrami, without any meridian cut, is a highly complicated problem, as is well known to shoes, ships and aircrafts makers \cite{CaiLiZhang}. Our goal was to ``squash'' it into a deformed annulus \cite{azariadisprivate}, without cutting the surface along a meridian (the truncation along some parallel on the ``tail'', instead, was required, as explained earlier). This would help retaining the necessary global properties of the Beltrami spacetime. Unfortunately, we realized that the strain would be too big \cite{azariadisprivate}, and would break the $\sigma$-bonds \cite{dominik}.

\begin{table}[h]
\caption{Estimates of radii of curvature as function of the number of defects, and corresponding threshold energies, for a 500 points lattice of molecular graphene. The last values correspond to include all the six heptagonal defects, required to mimic a Beltrami (and other) pseudosphere. The corresponding curvature is too close to the limit to have reliable results.}
\centering
\begin{tabular}{|l|l|c|}
  \hline
  $n_7^a$ & $r/\ell$ & $E_r$ (meV) \\ \hline
  1 & 21 & 10 \\ \hline
  2 & 15 & 14 \\ \hline
  3 & 12 & 17 \\ \hline
  4 & 10.5 & 20 \\ \hline
  5 & 9.5 & 22 \\ \hline
  6 & 8.5 & 25 \\
  \hline
\end{tabular}
\label{firsttable}
\end{table}

For all these reasons, the last proposal that was left to test the model, within these limits, was to see whether some relics of the effect are still there, for a configuration of planar molecular graphene that includes few heptagonal defects (with this, at least the condition of small curvature of the original surface is better satisfied/less unsatisfied, see table~\ref{firsttable}). In this fashion, the sample could mimic the planar development of a {\it portion} of a generic (not necessarily constant) negatively curved surface. The answer, as said, is negative. Let me explain why.

One piece of good news here arrived from the discovery that the issue (iii) of the list has positive answer. Namely, it is indeed possible to have a modified and approximate version of formula (\ref{THEformula}) at work, for small neighbors of a point of a graphene membrane shaped into a generic surface of constant negative curvature. Indeed, it is true that, as discussed in the $5^{\rm th}$ step of the previous Section (see \cite{ioriolambiase} and \cite{ioriolambiase2}), that the global predictability for the given surface in the lab frame is lost, but, as long as local considerations are concerned, some predictive power of (\ref{THEformula}) is still at work. Full details of this result will be given elsewhere. For the moment let me say that the main reason for this is that the horizon, in this model, is reached ``time-wise'', like in the standard Unruh effect, and not ``space-wise'', i.e. the horizon is not a specific curve on the manifold. Being the time $t$ the same for all surfaces in the lab frame $q^\mu$, the effect is still there for all, although modified and only valid portion by portion of the surface. 

Nonetheless (leaving aside the issue of constant vs nonconstant $\cal K$), one thing is to {\it focus} on one portion (neighbor of a point) of a whole surface, for which the global conditions for the existence of an horizon (hence the Hawking) are realized, and another thing is to have a {\it detached} (cut-away) portion of that surface. For instance, in order to feel the curvature, the electrons need ho have enough time to travel back and forth on the {\it whole} surface, an instance that also means that the electrons that feel the curvature are those of small energy. When this takes place on a surface, the electrons come back to the initial point (for a non-compact surface, at least some of them), while for a piece cut-away from the surface, obviously this cannot happen. Thus, there is no hope to see any effect that originates from global properties, as the Hawking-Unruh of (\ref{THEformula}).

To conclude on a positive note: although it was impossible to test these ideas in the experimental settings of Manoharan's lab (that were not meant to test these kind of effects, in the first place), the fact that we can test them on a generic surface of constant negative $\cal K$ is very promising for future, dedicated experiments.

\ack I thank P.~Azariadis, G.~Gibbons, and S.~Sen, for interesting correspondence, and G.~Lambiase and D.~Rastawicki, for collaboration on some of the matters discussed in Section 2 and in Section 3, respectively.

\end{document}